\documentclass[secnumarabic, graphics,floatfix, nofootinbib,tightenlines,nobibnotes, aps, prl, 12pt]{revtex4}
\usepackage{graphicx}
\usepackage{amsmath}

\begin{document}

\title{On the origin of trigger-angle dependence of di-hadron correlations}
\author{Wei-Liang Qian$^1$, Rone Andrade$^2$, Fernando Gardim$^2$, Fr\'ed\'erique Grassi$^2$, Yogiro Hama$^2$}
\affiliation{$^1$Departamento de F\'{\i}sica, Universidade Federal de Ouro Preto, MG, Brazil}
\affiliation{$^2$Instituto de F\'{\i}sica, Universidade de S\~ao Paulo, SP, Brazil}
\date{August 2, 2012}

\begin{abstract}
The STAR Collaboration reported measurements of di-hadron azimuthal correlation in medium-central Au+Au collisions at 200 A$\,$GeV, where the data are presented as a function of the trigger particle's azimuthal angle relative to the event plane $\phi_s\,$.
In particular, it is observed that the away-side correlation evolves from single- to double-peak structure with  increasing $\phi_s$. 
In this work, we present the calculated correlations as functions of both $\phi_s$ and particle transverse momentum $p_T\,$, using the hydrodynamic code NeXSPheRIO.
The results are found to be in reasonable agreement with the STAR data.
We further argue that the above $\phi_s$ dependence of the correlation structure can be understood in terms of one-tube model, as due to an interplay between the background elliptic flow caused by the initial state global geometry and the flow produced by fluctuations. 
\end{abstract}

\maketitle
\newpage

\section{I. Introduction}
Di-hadron correlations in ultrarelativistic heavy-ion collisions provide valuable information on the properties of the created medium.
The correlated hadron yields at intermediate and low $p_T$, 
when expressed in terms of the pseudo-rapidity difference $\Delta\eta$ and azimuthal angular spacing $\Delta\phi$, 
are strongly enhanced \cite{star-ridge1,star-ridge2,star-ridge3,phenix-ridge4,phenix-ridge5,phobos-ridge6,phobos-ridge7,alice-vn4,cms-ridge5,atlas-vn1} 
compared to those at high $p_T$ \cite{star-jet-1,star-jet-2}. 
The structure in the near side of the trigger particle is usually referred to as ``ridge''. It has a narrow $\Delta\phi$ located around zero and a long extension in $\Delta\eta$, and therefore it is tied to long range correlation in pseudo-rapidity. 
The away-side correlation broadens from peripheral to central collisions, and may exhibit double peak in $\Delta\phi$ for certain centralities and particle $p_T$ ranges. The latter is usually called ``shoulders''. These structures, for the most part, can be successfully interpreted as consequence of collective flow  \cite{sph-corr-1,sph-corr-3,hydro-v3-1,ph-vn-3,ph-corr-1,sph-corr-4} 
due to the hydrodynamical evolution of the system. 

Recently, efforts have been made to investigate the trigger-angle dependence of di-hadron correlations. 
STAR Collaboration reported measurements \cite{star-plane1} of di-hadron azimuthal correlations as a function of the trigger particle's azimuthal angle relative to the event plane, $\phi_s=|\phi_{tr}-\Psi_{EP}|$ at different trigger and associated transverse momenta $p_T$. The data are for 20-60\% mid-central 
Au+Au collisions at 200 A$\,$GeV. In a more recent study \cite{star-plane2},
the correlation was further separated into ``jet'' and ``ridge'', where the ridge yields are obtained by considering hadron pairs with large $|\Delta\eta|$. In this procedure, one assumes that the ridge is uniform in $\Delta\eta$ while jet yields are not. 
Such assumption is quite reasonable when one takes into account the measured correlation at low $p_T$ without a trigger particle \cite{star-ridge5}. In their work, all correlated particles at $|\Delta\eta| > 0.7$ are considered to be part of the ridge. 
It was observed that the correlations vary with $\phi_s$ in  both the near and away side of the trigger particle. 
The ``ridge'' drops when the trigger particle goes from in-plane to out-of-plane and, moreover, the correlations in the away-side evolve from single- to double-peak with increasing $\phi_s$. 

Owing to the $|\Delta\eta|$ cut, it is quite probable that these data mainly reflect the properties of the medium.
If this is the case, the main features of the observed 
trigger-angle dependence of di-hadron correlations should be  reproduced by hydrodynamic simulations, as hydrodynamic models have been shown capable of reproducing many important characteristics in collective flow  \cite{hydro-review-1,hydro-v2-heinz-1,hydro-v2-heinz-2,hydro-v2-heinz-3,hydro-v2-shuryak-1,hydro-v2-hirano-1,hydro-v2-hirano-2,sph-v2-1,sph-v2-2,hydro-v3-1,hydro-v3-3,hydro-eve-3,hydro-eve-4,epos-2}. 
The NeXSPheRIO code provides a good description of observed hadron spectra \cite{sph-cfo-1}, collective flow\cite{sph-v2-1,sph-v2-2,sph-v1-1,sph-vn-1}, elliptic flow fluctuations \cite{sph-v2-3}, and two-pion interferometry \cite{sph-hbt-1}.
In addition, it is known to reproduce the structures in di-hadron long-range correlations \cite{sph-corr-1}. In our previous studies on ridge \cite{sph-corr-3,sph-corr-2,sph-corr-5}, we obtained some of the experimentally known properties such as: centrality dependence and $p_T$ dependence. It is therefore interesting to see whether the model is further able to reproduce the observed trigger-angle dependence of the data. Moreover, it is intriguing to identify the underlying physical origin behind the phenomenon and numerical simulations. We, therefore, propose an intuitive explanation on the mechanism of the trigger-angle dependence of the ridge structures based on hydrodynamics. This is the main purpose of the present study. 

The paper is organized as follows. In the next section, we carry out a hydrodynamic study on the trigger-angle  dependence of di-hadron correlations by using NeXSPheRIO code. The calculations are done both with and without the pseudo-rapidity cut $|\Delta\eta| > 0.7\,$.  
Numerical results are presented for different angles of trigger particles and at different associated-particle transverse momentum, and they are compared with STAR data \cite{star-plane1,star-plane2}.
We try to understand the origin of the observed features in section III, by making use of an analytic parametrization of 
one-tube model \cite{sph-corr-3}. 
It is shown that the main features of the experimentally observed $\phi_s$ dependence can be obtained by the model,  where one takes into account the interplay of the flow harmonics caused by a peripheral energetic tube and those from bulk background.\footnote{A preliminary report of this discussion has been presented by Y.H. at the ISMD2011 meeting. \cite{sph-corr-ev-2}} In our approach, the background modulation is evaluated both by using cumulant and ZYAM method, and very similar results are obtained.
Section IV is devoted to discussions and conclusions.

\section{II. Numerical results of NeXSPheRIO}

Here, we present the numerical results on di-hadron azimuthal correlation, using the hydrodynamic code NeXSPheRIO. This code uses initial conditions (IC) provided by the event generator NeXuS \cite{nexus-1,nexus-rept},
solves the relativistic ideal hydrodynamic equations with SPheRIO code \cite{sph-review-1}. 
By generating many NeXuS events, and solving independently the equations of hydrodynamics for each of them, one takes into account the fluctuations of IC in event-by-event basis.
At the end of the hydrodynamic evolution of each event, a Monte-Carlo generator is employed to achieve hadron emission,
in Cooper-Frye prescription, and then hadron decay is considered. 

To evaluate di-hadron correlations, we generate 1200 NeXuS events in 20-60\% centrality window for 200 A$\,$GeV Au-Au collisions. At the end of each event, Monte-Carlo generator is invoked for decoupling, from 300 times for more central to 500 times for more peripheral collisions.  
Here we emphasize that there is no free parameter in the present simulation, since the few existing ones have been fixed in earlier studies of $\eta$ and $p_T$ distributions \cite{sph-v2-3}.   
To subtract the combinatorial background, we evaluate the two-particle cumulant. In order to make different events similar in characters, the whole centrality window is  further divided equally into four smaller centrality  classes, from 20-30\% to 50-60\%. Then one picks a trigger particle from one event and an associated particle from a different event to form a hadron pair. 
Averaging over all the pairs within the same sub-centrality class, one obtains the two particle cumulant. 
Background modulation is evaluated and subtraction is done within each sub-centrality class and then they are summed up together at the end. We calculated both cases: with and without the pseudo-rapidity cut. In the first case, all hadron pairs are included. Then in the latter case, only hadron pairs with $|\Delta\eta| > 0.7$ are considered as done in the STAR analysis. 
Note that in our approach, the IC constructed by ``thermalizing'' NeXus output do not explicitly involve jets, 
but they are not totally forgotten in the IC as they manifect themselves as high transverse fluid velocity in some localized region (see for instance Fig.2 of Ref.\cite{sph-corr-2}).
These regions with high transverse velocity, if not smeared out during hydrodynamic evolution, 
shall show up as a part of the near side ``jet" peak in the resulting two particle correlaiton.
Owing to its different physical origin, it is not correlated to the ``ridge'' and ``shoulder'' due to initial geometrical irregularities.
The implementation of the pseudo-rapidity cut in our calculation should reduce the correlation due to such effect.

The numerical results are shown in Fig.1 and Fig.2 in solid lines. They are, respectively, compared with the STAR data \cite{star-plane1,star-plane2} in filled circles and flow systematic uncertainties in histograms. From Fig.1 and Fig.2, one sees that the data are reasonably reproduced by NeXSPheRIO code. Correlations decrease both in the near side and in the away side when $\phi_s$ increases. 
For high-$p_T$ triggers, this was thought to be related to the path length that the parton traverses \cite{star-plane2}.
Here, we see that such feature is also presented by  intermediate- and low-energy particles, and reproduced well in a hydrodynamic approach. The magnitude of correlations in Fig.2 is globally smaller than those in Fig.1. 
This is because due to the $\Delta\eta$ cut in the former case, the total yields of associated particles is reduced.
If one only takes into consideration the overall shape of the correlations, one may notice that the hydrodynamic results in both plots are quite similar, Fig.2 can be approximately obtained if one scales the plots in Fig.1 by a factor of 0.6. The above results can be understood as a consequence of approximate Bjorken scaling in our hydrodynamic model. 
On one hand, the $\Delta\eta$ cut effectively removes a portion of associated particles with selected pseudo-rapidity difference to the trigger particle.
On the other hand, since the correlation is divided by the 
total trigger-particle number, the reduction of this does not affect the normalization. As expected, NeXSPheRIO results fit better for low momentum and the deviations increase at higher momentum. 
The calculated ridge structure in $\Delta\phi$ varies, especially on the away side, with trigger direction.
For in-plane trigger, simulation results exhibit a one-peak structure, which is broader in comparison with the near-side peak.
The away-side structure changes continuously from one peak at in-plane direction ($\phi_s=0$) 
to double peak at out-of-plane direction ($\phi_s=\pi/2$).
This characteristic is manifested particularly for larger transverse momentum of the associated particle.
All these features are in good agreement with the STAR measurements.

\section{III. The origin of trigger-angle dependence of ridge structure}

As pointed out, the motivation to introduce the $|\Delta\eta|$ cut was to separate ``ridge'' from ``jet'',
therefore the measured ``ridge'' correlations are expected to reflect the properties of the medium.
In the previous section, we showed that hydrodynamic simulations are able to reproduce the main features observed experimentally. 
Now, how this effect is produced?
In a previous study \cite{sph-corr-ev-1}, we tried to clarify the origin of the effect by using the one-tube model \cite{sph-corr-3} adapted to non-central collisions. Full details of the model can be found in \cite{sph-corr-3,sph-corr-2,sph-corr-4,sph-corr-5,sph-corr-6}. Here we only outline the main thoughts. In one-tube model, or more generally peripheral-tube model (allowing more than one tube), one treats the initial energy profile on a  transverse plane as superposition of peripheral high energy tubes on top of the background, which can be thought as the average distribution. The problem is further simplified by studying the transverse hydrodynamic expansion of a system consisting of only one peripheral tube with the assumption of longitudinal invariance. As the high energy tube deflects the collective flow generated by the background, the resulting single particle azimuthal distribution naturally possesses two peaks. They eventually give rise to the desired ``ridge'' and ``shoulder'' structures.
In ref.\cite{sph-corr-ev-1}, we took as background the average energy-density distribution obtained with NeXSPheRIO,
which has an elliptical shape for non-central collisions.
A peripheral tube sits on top of the background and its azimuthal position varies from event to event.
As shown there, the di-hadron correlation obtained from such a simple IC configuration does reproduce the main features of the data. 

Though the IC of the problem was greatly simplified through such an approach, the underlying physical mechanism of the obtained features were still not clear. In order to identify it more transparently, we further simplify it using an approximate analytical model. 
The derivation of the results relies on the following three hypotheses: 
\begin{itemize}
\item The collective flow consists of contributions from the background and those induced by a peripheral tube. 
\item A small portion of the flow is generated due to the interaction between background and the peripheral tube and, 
therefore, the flow produced in this process is correlated to the tube. 
\item Event by event multiplicity fluctuations are further considered as a correction that sits atop of the above collective flow of the system. 
\end{itemize} 
Let us comment briefly on these hypotheses. Based on the idea of one-tube model, a small portion of the background flow is deflected  by the peripheral tube; extra Fourier components of the flow are  generated by this process. The event plane of these extra flow harmonics are consequently correlated to the location of the tube, as stated in the second hypothesis. 
Since the contribution from the tube is small, we will treat it  perturbatively, considering the resultant flow a superposition of 
the background flow and the one produced in the tube-background 
interaction as described above. We believe that, at least  qualitative behavior of the results will remain valid also in more realistic case. Here, we are considering just one tube as fluctuation. However, in the limit of small perturbations, it is quite straightforward to generalize our results to the case of $N$ tubes. 
There are many possible sources for fluctuations, such as: flow fluctuations, multiplicity fluctuations etc, 
we will only consider multiplicity fluctuations in this simple model as assumed in the third hypothesis. 
As it will be shown below, this turn out to be enough to derive the observed feature in di-hadron correlations.

Using the hypotheses stated above, we write down the one-particle distribution as a sum of two terms: the distribution of the background and that of the tube. 
\begin{eqnarray} 
  \frac{dN}{d\phi}(\phi,\phi_t) =\frac{dN_{bgd}}{d\phi}(\phi) +\frac{dN_{tube}}{d\phi}(\phi,\phi_t), 
 \label{eq1}  
\end{eqnarray} 
where 
\begin{eqnarray} 
 \frac{dN_{bgd}}{d\phi}(\phi)&=&\frac{N_b}{2\pi}(1+2v_2^b\cos(2\phi)),    \label{eq2}\\  
 \frac{dN_{tube}}{d\phi}(\phi,\phi_t)&=&\frac{N_t}{2\pi}\sum_{n=2,3}2v_n^t\cos(n[\phi-\phi_t])   \label{eq3}     
\end{eqnarray} 
Since the background is dominated by the elliptic flow in non-central collisions, as observed experimentally, in Eq.(\ref{eq2}) we consider the most simple case, by parametrizing it with the elliptic flow parameter $v_2^b$ and the overall  multiplicity, denoted by $N_b$. As for the contributions from the tube, for simplicity, we assume they are independent of its angular position $\phi_t$ and
take into account the minimal number of Fourier components to reproduce the two-particle correlation due to the sole existence of a peripheral tube in an isotropic background (see the plots of Fig.2 of ref.\cite{sph-corr-4}, for instance). 
Therefore, only two essential components $v_2^t$ and $v_3^t$ are retained in Eq.(\ref{eq3}). We note here that the overall triangular flow in our approach is generated only by the tube, {\it i.e.}, $v_3=v_3^t$ and so its symmetry axis is correlated to the tube location $\phi_t$. 
The azimuthal angle $\phi$ of the emitted hadron and the position of the tube $\phi_t$ are measured with respect to the event plane $\Psi_2$ of the system. Since the flow components from the background are much bigger than those generated by the tube, as discussed below, $\Psi_2$ is essentially determined by the elliptic flow of the background $v_2^b$. 

The di-hadron correlation is given by
\begin{eqnarray} 
 \left<\frac{dN_{pair}}{d\Delta\phi}(\phi_s)\right>   =\left<\frac{dN_{pair}}{d\Delta\phi}(\phi_s)\right>^{proper} -\left<\frac{dN_{pair}}{d\Delta\phi}(\phi_s)\right>^{mixed} , 
 \label{eq4}  
\end{eqnarray} 
where $\phi_s$ is the trigger angle ($\phi_s=0$ for in-plane and $\phi_s=\pi/2$ for out-of-plane trigger). 
In one-tube model, 
\begin{eqnarray} 
 \left<\frac{dN_{pair}}{d\Delta\phi}\right>^{proper}  
 =\int\frac{d\phi_t}{2\pi}f(\phi_t)
 \frac{dN}{d\phi}(\phi_s,\phi_t)\frac{dN}{d\phi}  
 (\phi_s+\Delta\phi,\phi_t),   
\end{eqnarray}
where $f(\phi_t)$ is the distribution function of the tube. 
We will take $f(\phi_t)=1$, for simplicity. 

The combinatorial background 
$\left<{dN_{pair}}/{d\Delta\phi}\right>^{mixed}$
can be calculated by using either cumulant or ZYAM method \cite{zyam-1}. As shown below, both methods lend very similar conclusions in our model. 
Here, we first carry out the calculation using cumulant, which gives
\begin{eqnarray} 
\left<\frac{dN_{pair}}{d\Delta\phi}\right>^{mixed(cmlt)} 
 =\int\frac{d\phi_t}{2\pi}f(\phi_t)
 \int\frac{d\phi_t'}{2\pi}f(\phi_t')\frac{dN}{d\phi}   
 (\phi_s,\phi_t)\frac{dN}{d\phi}(\phi_s+\Delta\phi,\phi_t'). 
\end{eqnarray} 
Notice that, in the averaging procedure above, integrations both over $\phi_t$ and $\phi_t'$ are required in the mixed events, whereas only one integration over $\phi_t$ is enough for proper events. This will make an important difference between two terms in the subtraction of Eq.(\ref{eq4}). 

Using our simplified parametrization, Eqs.(\ref{eq1}-\ref{eq3}) and, by averaging over events, one obtains
\begin{eqnarray}
 \left<\frac{dN_{pair}}{d\Delta\phi}\right>^{proper} 
 &=&\frac{<N_b^2>}{(2\pi)^2}(1+2v_2^b\cos(2\phi_s))  
 (1+2v_2^b\cos(2(\Delta\phi+\phi_s))) \nonumber\\
 &+&(\frac{N_t}{2\pi})^2   
 \sum_{n=2,3}2(v_n^t)^2\cos(n\Delta\phi)
 \label{proper}
\end{eqnarray}
and       
\begin{eqnarray}       
 \left<\frac{dN_{pair}}{d\Delta\phi}\right>^{mixed(cmlt)}
 =\frac{<N_b>^2}{(2\pi)^2}(1+2v_2^b\cos(2\phi_s)
 (1+2v_2^b\cos(2(\Delta\phi+\phi_s))).
 \label{mixedc}                    
\end{eqnarray} 
Because of random distribution, contributions from peripheral tube 
are cancelled out upon averaging in the mixed-event correlation.   
Observe the difference between the factors multiplying the background terms of the proper- and mixed-event correlation. 
By subtracting Eq.(\ref{mixedc}) from Eq.(\ref{proper}), the resultant correlation is
\begin{eqnarray}
\left<\frac{dN_{pair}}{d\Delta\phi}(\phi_s)\right>^{(cmlt)} 
 &=&\frac{<N_b^2>-<N_b>^2}{(2\pi)^2}(1+2v_2^b\cos(2\phi_s))
 (1+2v_2^b\cos(2(\Delta\phi+\phi_s))) \nonumber\\  
 &+&(\frac{N_t}{2\pi})^2\sum_{n=2,3}2({v_n^t})^2
 \cos(n\Delta\phi). 
 \label{ccumulant}
\end{eqnarray} 
So, one sees that, {\it as the multiplicity fluctuates, the background elliptic flow does contribute to the correlation.}  
Now, by taking $\phi_s=0$ in Eq.(\ref{ccumulant}), the correlation for the in-plane trigger is finally given as 
\begin{eqnarray} 
 \left<\frac{dN_{pair}}{d\Delta\phi} \right>_{in-plane} ^{(cmlt)}
  &=&\frac{<N_b^2>-<N_b>^2}{(2\pi)^2} 
     (1+2v_2^b)(1+2v_2^b\cos(2\Delta\phi))\nonumber\\ 
  &+&(\frac{N_t}{2\pi})^2 
      \sum_{n=2,3}2(v_n^t)^2\cos(n\Delta\phi).  
 \label{eq9} 
\end{eqnarray} 
We note that, due to the summation of the two terms concerning $v_3^t$ and $v_2^b$, 
the away-side peak becomes broader than the near-side one, as shown below in Fig.\ref{fig3}. 

Similarly, the out-of-plane correlation is obtained by putting $\phi_s=\pi/2$ as 
\begin{eqnarray} 
 \left<\frac{dN_{pair}}{d\Delta\phi}\right>_{out-of-plane}^{(cmlt)}
  &=&\frac{<N_b^2>-<N_b>^2}{(2\pi)^2}
     (1-2v_2^b)
     (1-2v_2^b\cos(2\Delta\phi))\nonumber\\ 
  &+&(\frac{N_t}{2\pi})^2 
      \sum_{n=2,3}2(v_n^t)^2\cos(n\Delta\phi).   
 \label{eq10} 
\end{eqnarray} 
One sees that, because of the shift in the trigger angle $\phi_s$ ($0\rightarrow\pi/2$), the cosine dependence of the background contribution gives an opposite sign, as compared to the in-plane correlation. This negative sign leads to the following consequences. Firstly, there is a reduction in the amplitude of out-of-plane correlation both on the near side and on the away side. More importantly, it naturally gives rise to the observed double peak structure on the away side.
Therefore, despite its simplicity, the above analytic model reproduces the main characteristics of the observed data.
The overall correlation is found to decrease meanwhile away-side correlation evolves from a broad single peak to a double peak as $\phi_s$ increases. The correlations in Fig.3 is plotted with the following parameters
\begin{eqnarray} 
<N_t^2>=0.45 \nonumber \\
v_2^t=v_3^t=0.1 \nonumber \\
<N_b^2>-<N_b>^2=0.022 \nonumber \\
v_2^b=0.25  \label{parameterset}
\end{eqnarray} 
We note that both the correlated yields and the flow harmonics 
are actually dependent on the specific choice of the $p_T$ interval of the observed hadrons as shown in Fig.1 and Fig.2.
Since the transverse momentum dependence has not been explicitly taken into account in this simple model,
the multiplicities in the above parameter set are only determined up to an overall normalization factor, 
and the flow coefficients are chosen to reproduce the qualitative behavior of the trigger-angle dependence of di-hadron correlation as shown by data.


Now we will show that very similar results will be again obtained, 
if one evaluates the combinatorial mixed event contribution using 
ZYAM method. The spirit of ZYAM method is to first estimate the 
form of resultant correlation solely due to the average background collective flow and, then, the evaluated correlation is rescaled by 
a factor $B$, the latter is determined by assuming zero signal at minimum of the subtracted correlation. 
Di-hadron correlation for the background flow is given by 
\begin{eqnarray}       
 \left<\frac{dN_{pair}}{d\Delta\phi}\right>^{mixed(ZYAM)}
  =B(\phi_s)\int\frac{d\phi}{2\pi}\delta(\phi-\phi_s)
 \frac{dN_{bgd}}{d\phi}(\phi)\frac{dN_{bgd}}{d\phi}(\phi+\Delta\phi).
 \label{mixedz1}                    
\end{eqnarray}
In the STAR analyses, both $v_2$ and $v_4$ have been taken into account for the background. In our simplified model, however, the average 
background flow contains, besides the radial one, only the elliptic  flow $v_2^b$. Therefore, a straightforward calculation gives
\begin{eqnarray}       
 \left<\frac{dN_{pair}}{d\Delta\phi}\right>^{mixed(ZYAM)}
  =B(\phi_s)\frac{<N_b>^2}{(2\pi)^2}(1+2v_2^b\cos(2\phi_s)(1+2v_2^b\cos(\Delta\phi+\phi_s)).
 \label{mixedz2}                    
\end{eqnarray} 
We remark that, since in ZYAM method fluctuations are not explicitly 
considered, the multiplicity of background distribution, as given 
by Eq.(\ref{eq2}), is evidently average multiplicity $<N_b>$. 
By combining this term with the proper correlation, given by 
Eq.(\ref{proper}), the resultant correlation reads
\begin{eqnarray}
 \left<\frac{dN_{pair}}{d\Delta\phi}(\phi_s)\right>^{(ZYAM)} 
 &=&\frac{<N_b^2>-B(\phi_s)<N_b>^2}{(2\pi)^2}(1+2v_2^b\cos(2\phi_s))(1+2v_2^b\cos(2(\Delta\phi+\phi_s))) \nonumber\\
 &+&(\frac{N_t}{2\pi})^2\sum_{n=2,3}2{v_n^t}^2\cos(n\Delta\phi) \label{czyam} 
\end{eqnarray} 
The consistency of this expression with what is used in the STAR analyses will be discussed below, in the next section.   
Note that the normalization factor $B(\phi_s)$ is a function of trigger angle. 
It is fixed to give zero yield at the minimum of the subtracted correlation, namely $\left<{dN_{pair}}/{d\Delta\phi}(\phi_s)\right>^{(ZYAM)}=0$ at the minimum. 
Because the correlation in Eq.(\ref{czyam}) is positively defined, and the second term can be positive or negative,
the coefficient of the first term $<N_b^2>-B(\phi_s)<N_b>^2$ must be positive. 
One sees clearly that the above expression is almost identical to 
the cumulant result, Eq.(\ref{ccumulant}), so will cause similar 
trigger-angle dependence. 
This can also be seen from the plots of correlations obtained by adopting the same parameters as in Eq.(\ref{parameterset}) and additionally
\begin{eqnarray} 
<N_b^2>=100 , 
\end{eqnarray} 
where the value of $<N_b^2>$ is choosen to be larger compared to its fluctuation. 
The scale factor $B(\phi_s)$ is subsequently fixed by the minimum condition as 
\begin{eqnarray} 
B(\phi_s=0) = 1.000053 \nonumber \\
B(\phi_s=\pi/2)= 1
\end{eqnarray} 
The in-plane correlation plot is shown in Fig.4, which is very close to the one in Fig.3. The one-tube contribution and out-of-plane  correlation are not plotted, since they are exactly the same as those shown in Fig.3.


\section{IV. Discussions and Conclusions}

Here, we first show that the expression of di-hadron correlation 
of background flow in Eq.(\ref{mixedz2}) is in agreement with that obtained in Ref.\cite{ph-corr-ev-1}, 
which is employed in STAR analyses \cite{star-plane1,star-plane2}. The only difference is that we have only taken into account the second-order harmonic, and the reason for not including any higher-order flow components in our calculation is simply because we wanted to transparently show the  mechanism of in-plane/out-of-plane effect by using a model as simple as possible. 
In Ref.\cite{ph-corr-ev-1}, it was shown that
\begin{equation}
\frac{dN_{pair}}{d\Delta\phi}=B^{(R)}\left[1+2\sum_{n=1}^{\infty}v_n^{(a)}v_n^{(t,R)}\cos(n\Delta\phi)\right],\label{eq:bkgd}
\end{equation}
where $v_n^{(a)}$ is the associated particle's $n$-th harmonic, $v_n^{(t,R)}$ is the average $n$-th harmonic of the trigger  particles, and $B^{(R)}$, the background normalization, denotes the integrated inclusive pair yield
\begin{eqnarray} 
B^{(R)}&=&B\left(1+\sum_{k=2,4,6,...}2v_k^{(t)}\cos(k\phi_s)\frac{\sin(kc)}{kc}\right), \nonumber \\
v_n^{(t,R)}&=&\frac{v_n^{(t)}+\delta_{n,{\rm even}}T_n+\sum_{k=2,4,6,...}\left(v_{k+n}^{\rm (t)}+v_{|k-n|}^{\rm (t)}\right)T_k}
{1+\sum_{k=2,4,6,...}2v_{k}^{(t)}T_k}\ , \label{eq:vR} \\
T_k&=&\cos(k\phi_s)\frac{\sin(kc)}{kc}\left< \cos(k\Delta\Psi)\right>\,, \nonumber 
\end{eqnarray} 
where $2c$ is the angular width where a trigger is located. 
In our approach, the size of the slice is taken to be infinitely small ($c\to 0$), and perfect event plane resolution 
($\left< \cos(k\Delta\Psi)\right>=1$) is assumed. 
Take, for instance, the in-plane correlation by substituting 
$\phi_s =0$, and considering terms up to the second order, one 
obtains
\begin{eqnarray} 
B^{(R)}&=&B\left(1+2v_2^{(t)}\right) \nonumber \\
v_2^{(t,R)}&=&1 \nonumber 
\end{eqnarray} 
which is readily shown to be consistent with Eq.(\ref{mixedz2}).
In fact, it is intuitive to understand since, the one-particle distribution of the trigger in this case is a $\delta$ function which peaks at $\phi_s=0$.

In our approach, the trigger-angle dependence of di-hadron correlation is understood as due to the interplay between the 
elliptic flow  caused by the initial almond deformation of the whole system and flow produced by fluctuations. 
The contributions due to fluctuations are expressed in terms of a high-energy-density tube and the flow deflected by it. 
However, the generic correlation due to the tube is preserved even after the background subtraction\cite{sph-corr-4}, 
by this simple model we show explicitly that the result does not depend on either cumulant or ZYAM method.
This is because the form of combinatorial background is determined by the average flow harmonics,
as shown in Eqs.(\ref{ccumulant}) and (\ref{czyam}).
Though in this approach, only elliptic flow is considered for simplicity, it is straightforward to extend the result here to a more general case.
Due to multiplicity fluctuations or due to the procedure in ZYAM, the background flow may also contribute to the subtracted correlation.
Since the background modulation is shifted, changing the phase,  when the trigger particle moves from in-plane to out-of-plane as seen in Eqs(\ref{mixedc}) and (\ref{mixedz2}),
the summation of the contributions of the background and that of the tube give rise to the desired trigger-angle dependence.

In the one-tube model, a part of the flow is caused by the peripheral energetic tube. 
Since the tube deflects the global flow of the background, 
the event planes of such flow components are correlated to the localization of the tube as expressed in Eq.(\ref{eq3}).
Particularly, it also contains the second harmonic $v_2^t$.
Though it is present in the proper two particle correlation (Eq.(\ref{proper})), it is not considered in Eq.(\ref{mixedz2}) when evaluating combinatorial background correlation.
The reason is two-fold. 
Firstly, to calculate the average $v_2$ of a given event, one must use multiplicity as weight, in our model, 
the multiplicity of background $N_b$ is assumed to be much bigger than that of the tube $N_t$.
(The parameter $<N_b^2>=100$ can be freely changed to a much bigger number.)
Moreover, since $\phi_t$ varies from event to event, 
the contribution to $v_2$ from $v_2^t$ is positive at $\phi_t=0$ and negative at $\phi_t=\pi/2$.
When averaging over different events, most contributions cancel 
each other at different $\phi_t$ values.
As it happens, $v_2^t$ contributes to the subtracted correlation while it does not manifest itself in average background flow.
This is an important feature of the present approach.

It is interesting to note that the two particle correlation 
has also been studied using Fourier expansion in\cite{ph-corr-1,ph-corr-ev-2}
\begin{eqnarray}
 \left<\frac{dN_{pair}}{d\Delta\phi}(\phi_s)\right>^{proper} 
 &=&\frac{N^2}{(2\pi)^2}(1+2V_{2\Delta}\cos(2\Delta\phi)+2V_{3\Delta}\cos(3\Delta\phi))+\cdots
\end{eqnarray}
For comparison, we rewrite Eq.(\ref{proper}) in terms of $V_{n\Delta}$ as follows
\begin{eqnarray}
V_{2\Delta}&=&\frac{N_t^2}{\left<N_b^2\right>\left(1+ 2v_2^b\cos(2\phi_s)\right)}\left(v_2^t\right)^2+ \cos(2\phi_s)v_2^b \label{V2d} \\
V_{3\Delta}&=&\frac{N_t^2}{\left<N_b^2\right>\left(1+ 2v_2^b\cos(2\phi_s)\right)}\left(v_3^t\right)^2 \label{V3d}\, .
\end{eqnarray}
One sees that the background elliptic flow $v_2^b$ dominates $V_{2\Delta}$ for both in-plane and out-of-plane directions, 
while $V_{3\Delta}$ is determined by the triangular flow $v_3^t$ produced by the tube.
Due to the factor $\cos(2\phi_s)$, the second term of Eq.(\ref{V2d}) changes sign when the trigger angle goes from $\phi_s=0$ to $\phi_s=\pi/2$.
Dominated by this term, $V_{2\Delta}$ decreases with $\phi_s$, and it intersects $V_{2\Delta}=0$ at around $\phi_s=\pi/4$.
Since the first term in Eq.(\ref{V2d}) is positive definite, the integral of $V_{2\Delta}$ with respect to $\phi_s$ is positive.
These features are in good agreement with the data analysis (see Fig.1 of ref.\cite{ph-corr-1}).
On the other hand, the axis of triangularity is determined by the position of the tube $\Psi_3=\phi_t$.
Since we have assumed uniform distribution $f(\phi_t)=1$ in our calculation, 
the event plane of triangularity is totally uncorrelated with the event plane $\Psi_2$, as generally understood \cite{hydro-v3-1,glauber-en-3},
and consequently, the contribution from triangular flow should not depend much on the event-plane angle.
This is indeed shown in the above expression Eq(\ref{V3d}).
$V_{3\Delta}$ barely depends on $\phi_s$, if anything, it slightly increases with increasing $\phi_s$.
This characteristic is also found in the data\cite{ph-corr-1}.

In conclusion, the NeXSPheRIO code gives correct qualitative behavior of the in-plane/out-of-plane effect. 
Physically, we understand that
this effect appears because, besides the contribution coming from the peripheral tube,
additional contribution from the background flow has to be considered. 
The latter is back-to-back (peaks at $\Delta\phi= 0 , \pi$) in the case of in-plane triggers ($\phi_s=0$) 
and rotated by $\pi/2$ (peaks at $\Delta\phi=-\pi/2 , \pi/2$) in the case of out-of-plane triggers ($\phi_s = \pi/2$).
A simplified analytical model is proposed, and it is shown to successfully reproduce the observed features.

\section{V. Acknowledgments}
We thank for valuable discussions with Fuqiang Wang, Paul Sorensen, Lanny Ray, Roy Lacey, Jiangyong Jia and Wei Li.
We acknowledge funding from Funda\c{c}\~ao de Amparo \`a Pesquisa 
do Estado de S\~ao Paulo (FAPESP), Funda\c{c}\~ao de Amparo \`a  Pesquisa do Estado de Minas Gerais (FAPEMIG) and Conselho Nacional  de Desenvolvimento Cientit\'{\i}fico e Tecnol\'ogico (CNPq).

\bibliographystyle{h-physrev}
\bibliography{references_qian}{}

\begin{figure}
  \centerline{\includegraphics[width=16.cm]{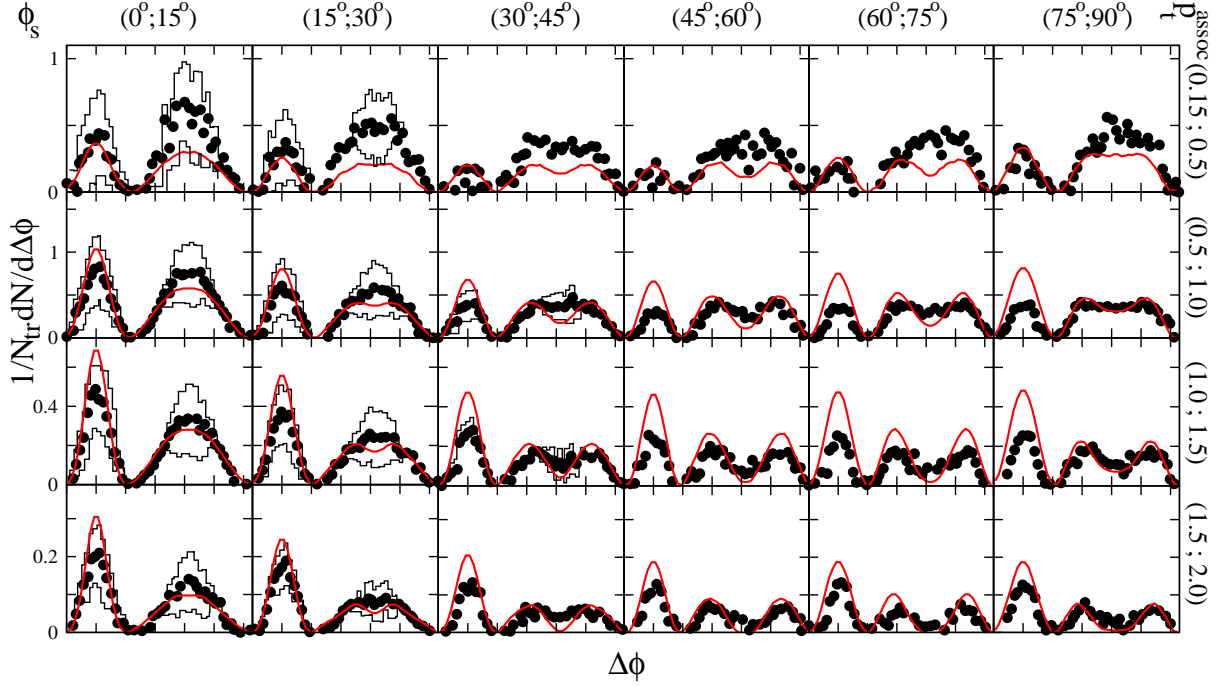}}
  \caption{
   The subtracted di-hadron correlations as a function of 
   $\Delta\phi$ for different $\phi_s=\phi_{trig}-\phi_{EP}$ and 
   $p_T^{assoc}$ with $3 < p_T^{trig} < 4 GeV$ in 20 - 60\% 
   Au+Au collisions. The $\phi_s$ range increases from 
   0-15$^{\circ}$ (left column) to 75-90$^{\circ}$ (right column); 
   the $p_T^{assoc}$ range increases from 0.15-0.5 GeV (top row) to 
   1.5-2 GeV (bottom row). NeXSPheRIO results in solid curves, are 
   compared with STAR data in filled circles~\cite{star-plane1}. 
   The histograms indicate  the systematic uncertainties from flow 
   subtraction. 
   } 
  \label{fig1}
\end{figure}

\begin{figure}
  \centerline{\includegraphics[width=16.cm]{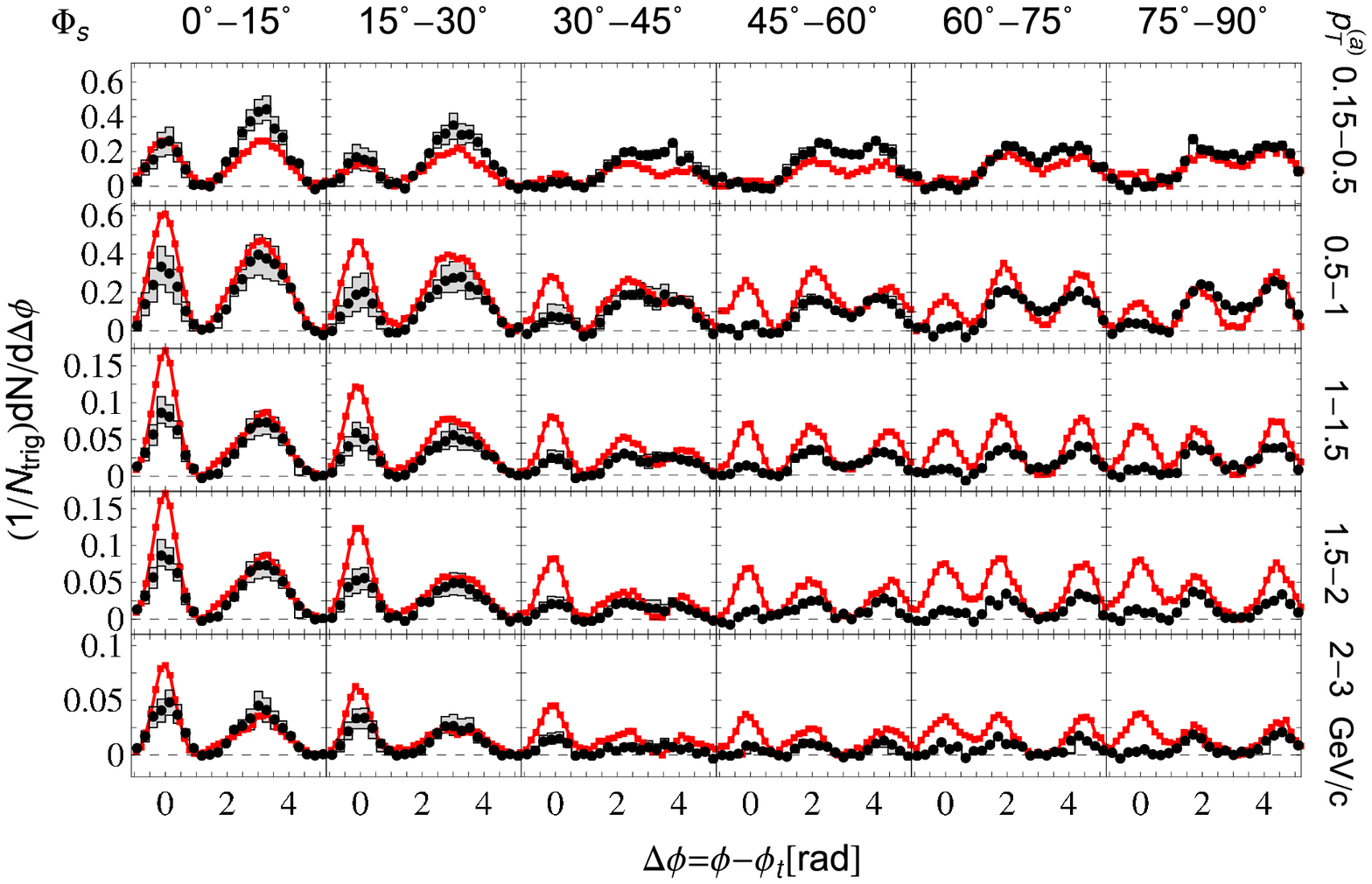}}
  \caption{
   The subtracted di-hadron correlations as a function of 
   $\Delta\phi$ for different $\phi_s=\phi_{trig}-\phi_{EP}$ and 
   $p_T^{assoc}$ with $3 < p_T^{trig} < 4 GeV$ and 
   $|\Delta\eta| > 0.7$ in 20 - 60\% Au+Au collisions. 
   The $\phi_s$ range increases from 0-15$^{\circ}$ (left column) 
   to 75-90$^{\circ}$ (right column); the $p_T^{assoc}$ range 
   increases from 0.15-0.5 GeV (top row) to 2-3 GeV (bottom row). 
   NeXSPheRIO results in solid curves are compared with STAR data 
   in filled circles~\cite{star-plane2}. The grey histograms 
   indicate the systematic uncertainties from flow subtraction.
   }
  \label{fig2}
\end{figure}

\begin{figure}
\begin{tabular}{cc}
\begin{minipage}{200pt}
\centerline{\includegraphics[width=220pt]{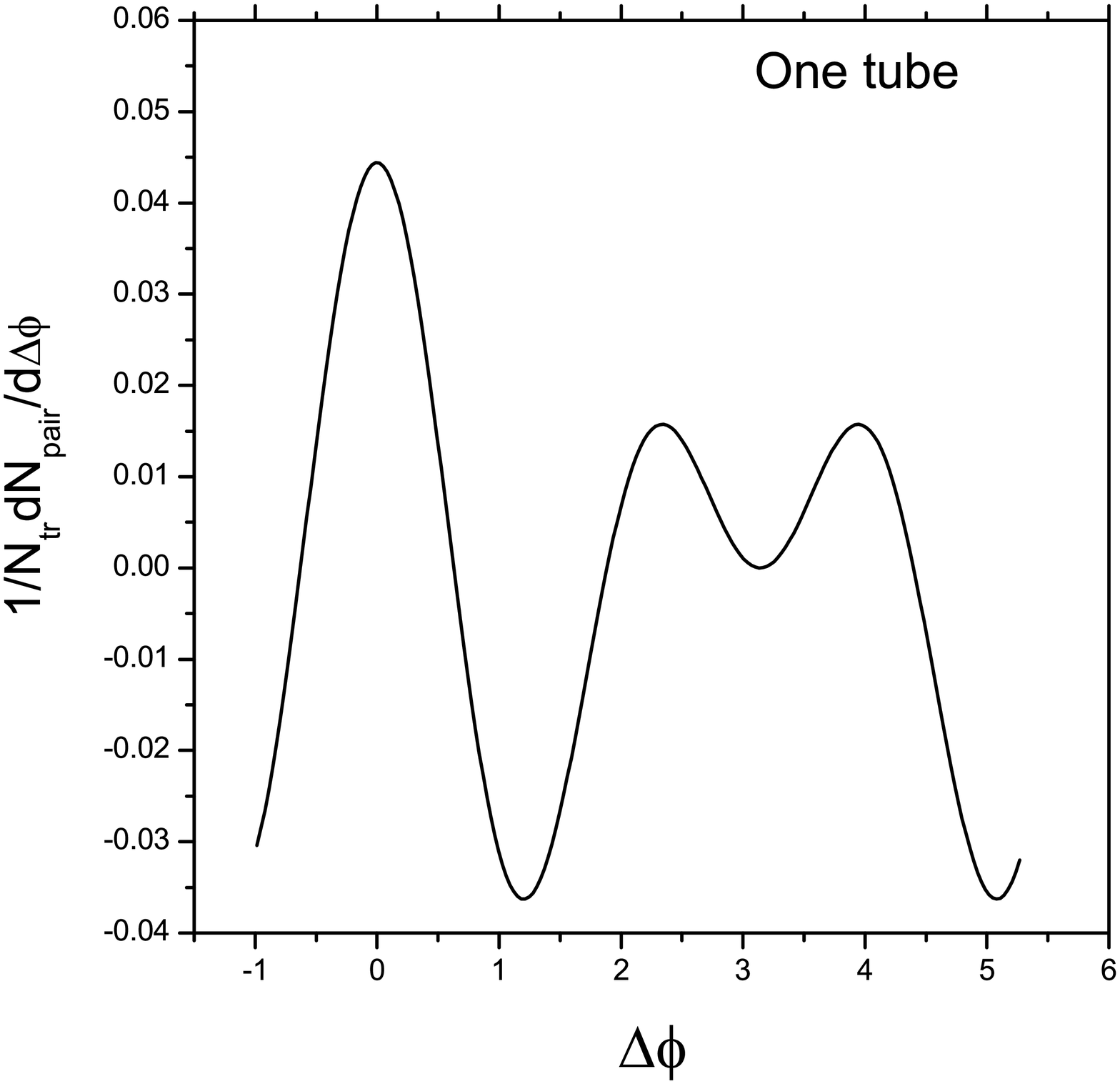}} 
\end{minipage}
&
\begin{minipage}{200pt}
\centerline{\includegraphics[width=220pt]{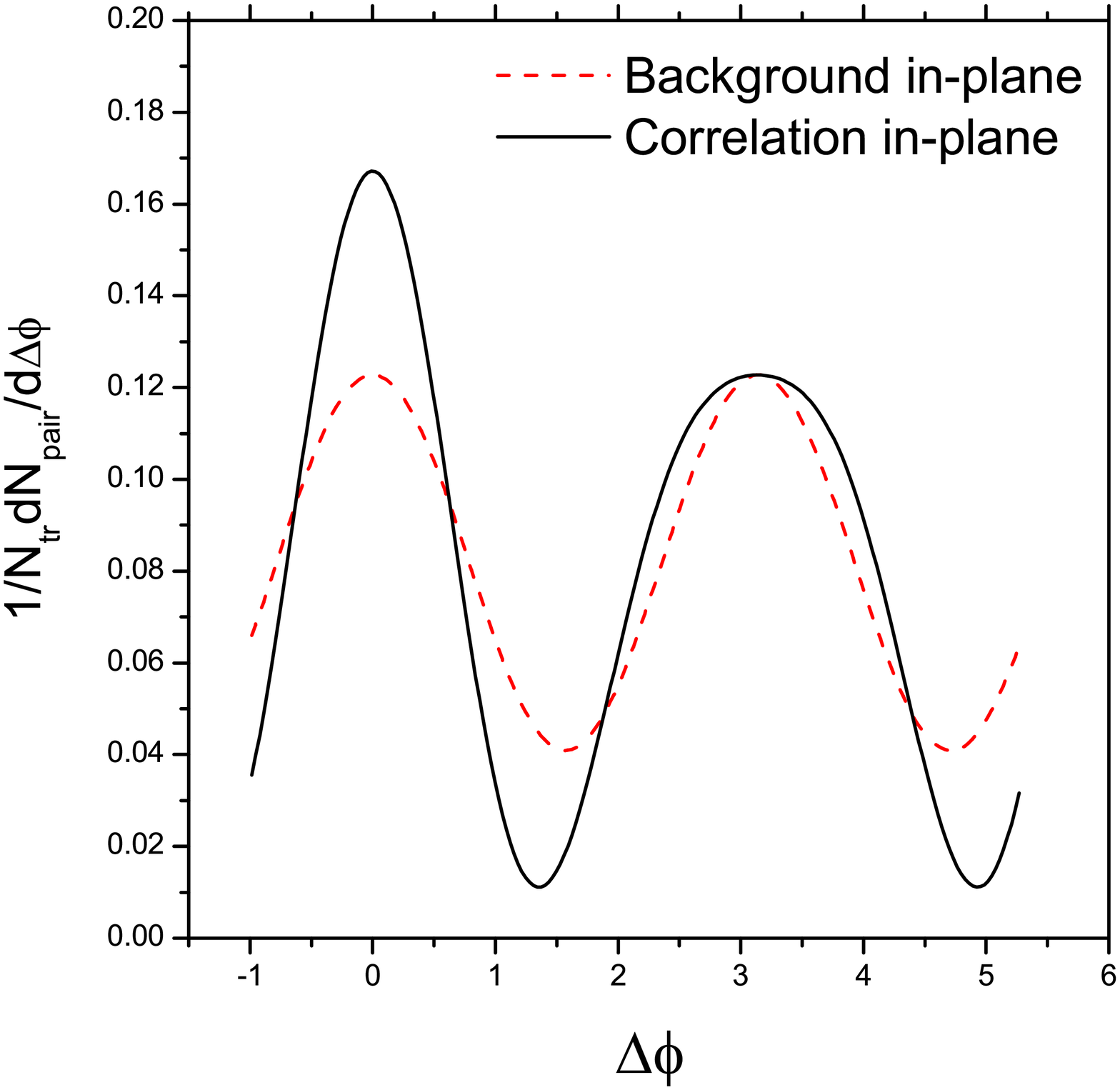}}
\end{minipage}
\\
\end{tabular}
\begin{minipage}{200pt}
\centerline{\includegraphics[width=220pt]{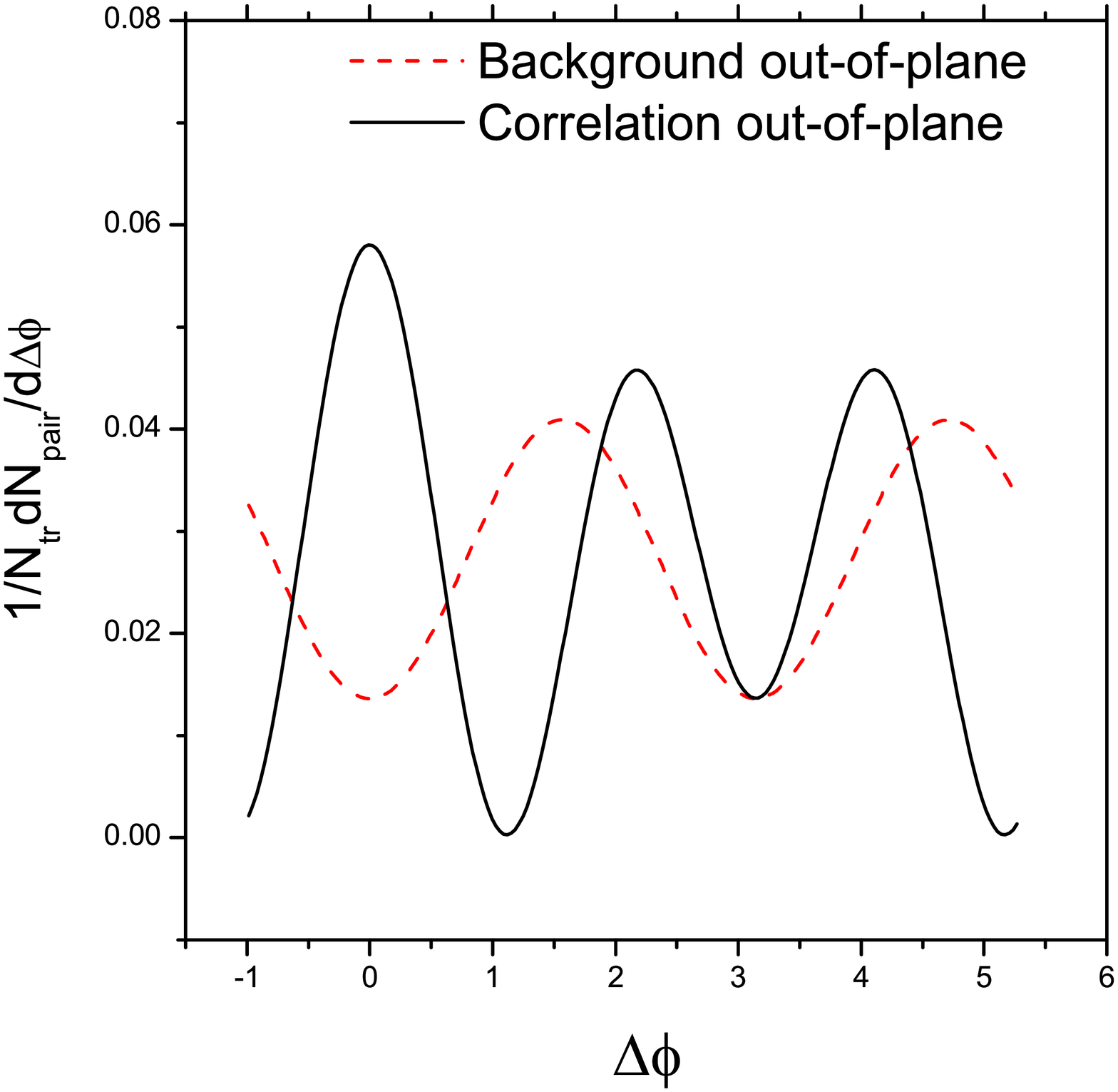}} 
\end{minipage}
 \caption{Plots of di-hadron correlations calculated by cumulant 
 method. From the left to the right: (i) the peripheral-tube 
 contribution; (ii) the one from the background (dashed line) and 
 the resultant correlation (solid line) for in-plane triggers, as 
 given by Eq.(\ref{eq9}); and (iii) the corresponding ones for the 
 out-of-plane triggers, Eq.(\ref{eq10}).} 
 \label{fig3}           
\end{figure} 

\begin{figure}
\centerline{\includegraphics[width=220pt]{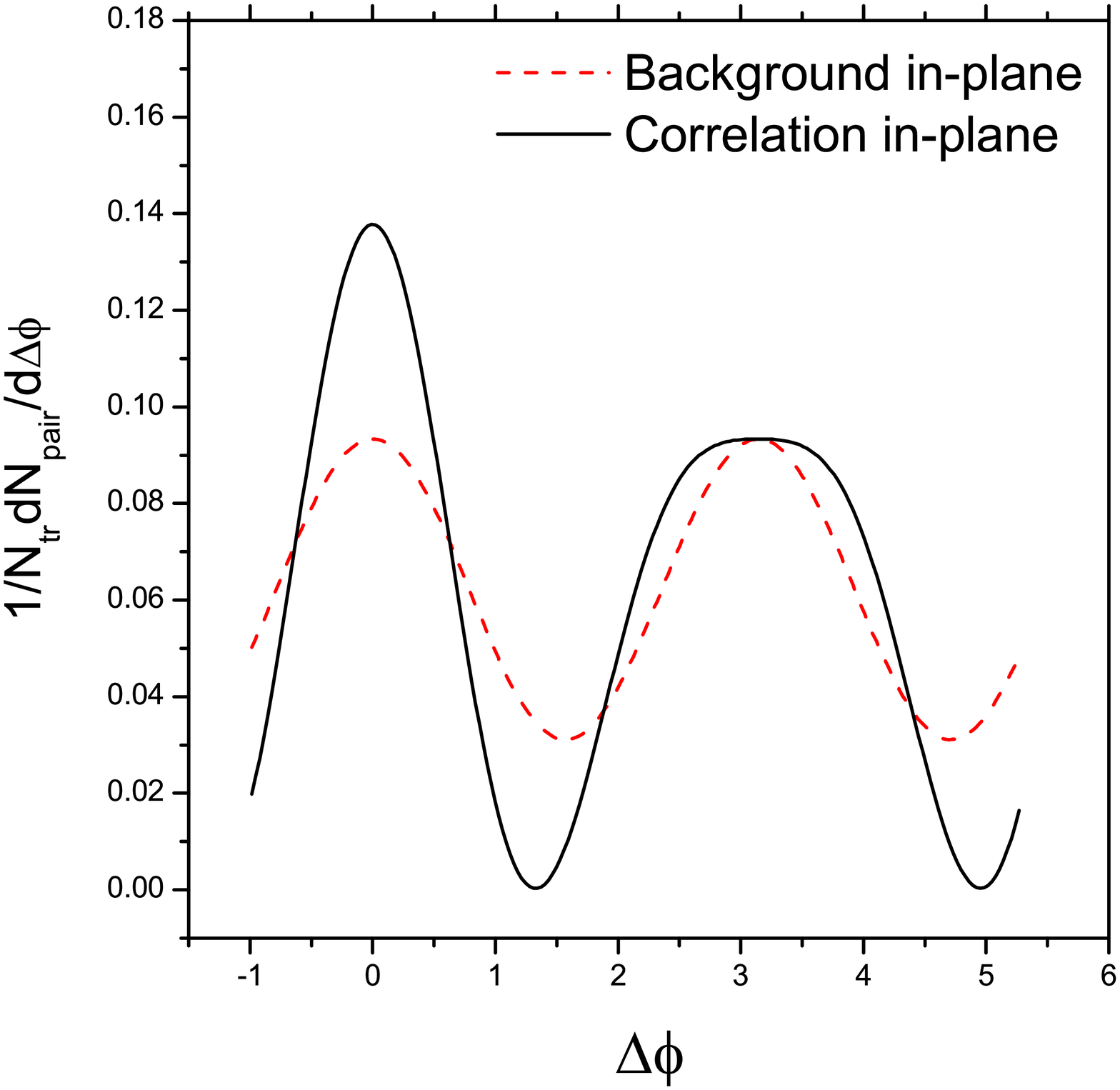}} 
 \caption{Plot of in-plane di-hadron correlation by using ZYAM as 
 given by Eq.(\ref{czyam}). The one-tube contribution and 
 out-of-plane correlation are exactly the same as the ones shown in 
 Fig.3.} 
 \label{fig4}           
\end{figure} 

\end{document}